\documentclass[aps,prb,amsmath,amssymb,twocolumn,superscriptaddress,floatfix]{revtex4-1}
\usepackage{graphicx}
\usepackage{hyperref}
\hypersetup{pdftitle={Al4W and Friends},pdfauthor={Darren C. Peets},pdfsubject={Al4W},pdfdisplaydoctitle}

\def\four{Al\ensuremath{_4}W}
\def\five{Al\ensuremath{_5}W}
\def\fiveMo{Al\ensuremath{_5}Mo}
\def\ttMo{Al\ensuremath{_{22}}Mo\ensuremath{_{5}}}
\def\fnMo{Al\ensuremath{_{49}}Mo\ensuremath{_{11}}}
\def\three{Al\ensuremath{_{3.808(37)}}W}

\def\twelve{Al\ensuremath{_{12}}Mo}
\def\Tc{\ensuremath{T_\text{c}}}
\begin{document}

\title{Physical properties of noncentrosymmetric tungsten and molybdenum aluminides}

\author{Darren C. Peets}
\email{dpeets@fudan.edu.cn}
\author{Tianping Ying}
\author{Xiaoping Shen}
\author{Yunjie Yu}
\affiliation{State Key Laboratory of Surface Physics, Department of Physics; and Advanced Materials Laboratory, Fudan University, Shanghai 200438, China}

\author{Maxim Avdeev}
\affiliation{Australian Nuclear Science and Technology Organisation, Lucas Heights, NSW 2234, Australia}

\author{Shiyan Li}
\author{Donglai Feng}
\affiliation{State Key Laboratory of Surface Physics, Department of Physics; and Advanced Materials Laboratory, Fudan University, Shanghai 200438, China}
\affiliation{Collaborative Innovation Center of Advanced Microstructures, Nanjing 210093, China}

\begin{abstract}

  A lack of spatial inversion symmetry gives rise to a variety of unconventional physics, from noncollinear order and Skyrmion lattice phases in magnetic materials to topologically-protected surface states in certain band insulators, to mixed-parity pairing states in superconductors.  The search for exotic physics in such materials is largely limited by a lack of candidate materials, and often by difficulty in obtaining crystals.  Here, we report the single crystal growth and physical properties of the noncentrosymmetric tungsten aluminide cage compounds \four\ and \five, alongside related molybdenum aluminides in which spin-orbit coupling should be significantly weaker.  All compounds are nonmagnetic metals.  Their high conductivities suggest the opportunity to find superconductivity at lower temperatures, while the limits we can place on their transition temperatures suggest that any superconductivity may be expected to exhibit significant parity mixing.

\end{abstract}

\maketitle

\section{Introduction}

Spatial inversion symmetry is sufficiently common in the crystal structures of materials that the constraints it places on electronic wavefunctions, notably parity, often underpin our understanding of physics.  In materials that lack spatial inversion symmetry, a variety of additional terms can be present in the Hamiltonian, such as Dzyaloshinskii-Moriya terms in magnets\cite{Dzyaloshinsky1958,Moriya1960}, Rashba-Dresselhaus spin-orbit band splitting\cite{Rashba1960,Dresselhaus1955}, or additional Coulomb terms\cite{Park2012,Kim2012,Kim2013,Park2015}.  These compete with existing interactions, and can lead to a rich array of entirely new physics either on their own or through that competition.  Examples include noncollinear magnetism, Skyrmions, and spin-split band structure.  The role of inversion is particularly clear in superconductors, where the orbital component of the pairing function usually inherits parity from the electron wavefunctions, then the spin component must be singlet or triplet to maintain Pauli exclusion.  An absence of spatial inversion means that parity is no longer meaningful, and any superconducting condensate is expected to be a mixture of singlet and triplet components, leading to a wide variety of unusual properties\cite{bauerbook,Fujimoto2007,Smidman2017}.  However, realizing these predictions has proven challenging.  


Prior to the discovery of the bismuthate and cuprate superconductors, the search for new superconducting materials was guided by a set of rules developed by Bernd Matthias\cite{Matthias1955,MatthiasRules}.  These rules, for instance, suggest that oxygen and magnetism be avoided, while certain electron fillings and high crystal symmetry are preferred.  The latter rule is broken in noncentrosymmetric superconductors.  Few such superconductors are known, perhaps in part due to decades spent following rules which disfavor them, and many of those that are known have not been prepared in single-crystalline form.  In our quest to identify more such superconductors by finding noncentrosymmetric metals of which crystals can be grown, we identified \four\ as potentially interesting.  This cage compound crystallizes in the monoclinic space group $Cm$ (No.\ 8)\cite{Bland1958}, while tungsten's position low on the periodic table may introduce the spin-orbit coupling required to produce spin-split band structure.  In fact, \four\ is just one member of a family of transition metal aluminides, most of which are noncentrosymmetric, in which the transition metal atoms are contained in 10- to 12-membered Al cages.


Here, we report the low-temperature properties of several related noncentrosymmetric tungsten and molybdenum aluminides.  All are excellent metals, but we are only able to put upper limits on any possible transition temperatures.  That single-crystalline intermetallic metals composed predominantly of aluminum do not superconduct down to in some cases 100\,mK is surprising, and may point to difficulties in forming pairs in the presence of spin-dependent band splitting or, more generally, strong spin-orbit coupling.

\section{Experimental}

\begin{figure}[htb]
  \includegraphics[width=\columnwidth]{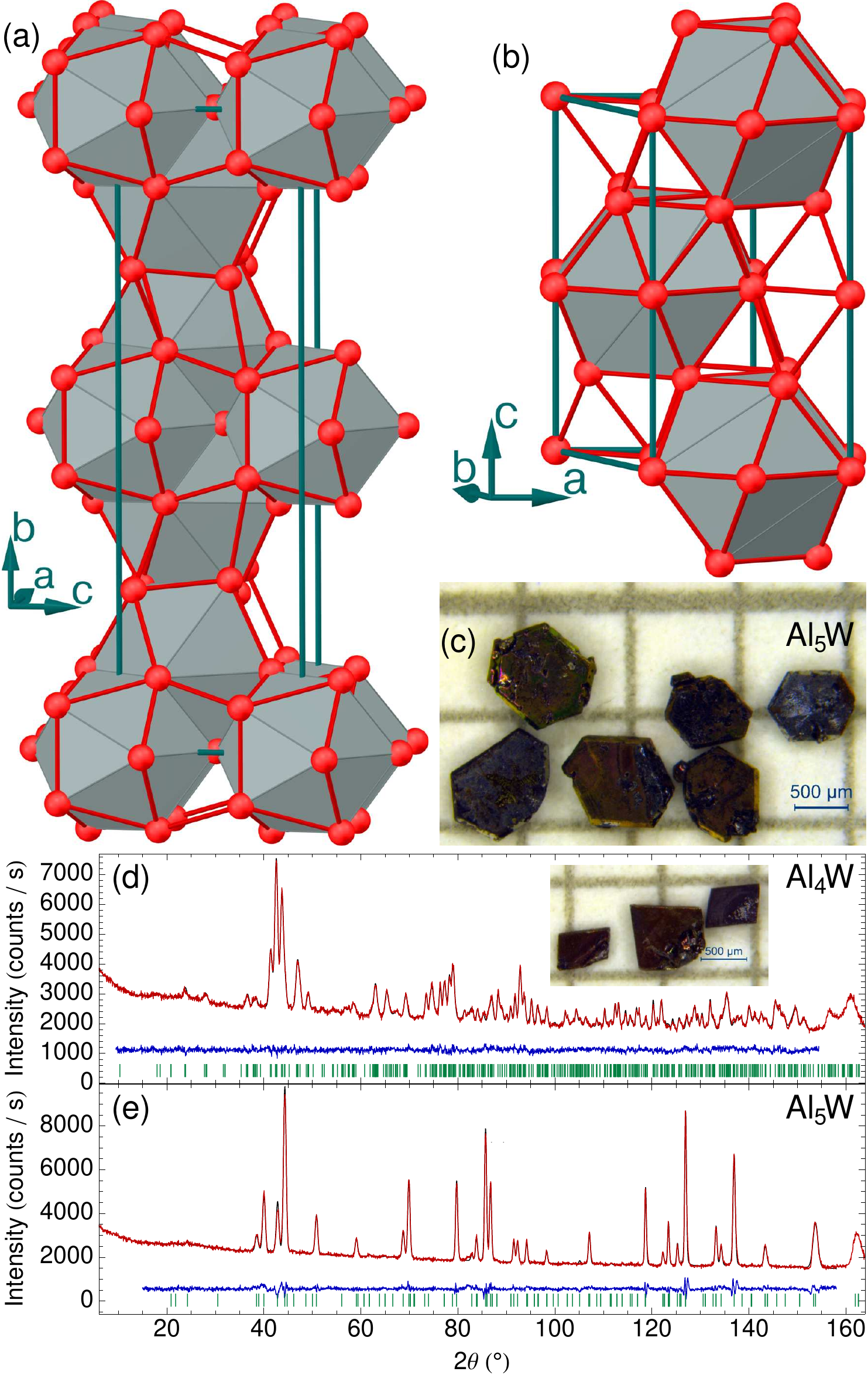}
  \caption{\label{fig:NPD}Refined crystal structures of (a) \four\ and (b) \five, with Al atoms in red and W-centred polyhedra in gray.  Al--Al bonds are shown for bond lengths less than 2.9\,\AA.  Examples of some of the \five\ crystals grown are shown in panel (c), with \four\ crystals appearing in the inset of panel (d).  Neutron powder refinements are shown in panels (d) and (e) for \four\ and \five, respectively.  Data are shown in red, the refinements in black, residuals in blue, and Bragg positions in green.}
\end{figure}

\four\ and \five\ melt peritectically at 1327 and 871\,$^\circ$C, respectively, but will crystallize out of Al flux\cite{Tonejc1972}.  Al wire (PRMat, 99.999\%) and tungsten powder (Alfa Aesar, 99.95\%), in the approximate ratio 60:1, were weighed into alumina crucibles inside an Ar-filled glovebox, and these crucibles were then sealed under vacuum inside quartz tubes.  The solubility of W in Al is extremely low at the relevant temperatures, so the stoichiometry was based more on the desired volume of crystals than on the location of the liquidus curve.  Alumina lids were used on \four, since both liquid and gaseous Al attack quartz and the vapor pressure of Al becomes a concern at the higher temperatures required.  To produce \four, the crucible was heated to 1050\,$^\circ$C at 200\,$^\circ$C/h, held at that temperature for 2\,h, then cooled to 900\,$^\circ$C over the course of 3--7 days.  At this point, the furnace was allowed to cool freely to room temperature.  For \five, a maximum temperature of 850\,$^\circ$C was used, and the growths ended at 720\,$^\circ$C.  Finally, the Al flux was dissolved off in 1\,M HCl, revealing mm$^2$-size platelet single crystals up to 300\,$\mu$m thick.  In some cases a thin dark film, most likely metal chlorides and chloride hydrates, had to be removed from the surface after this step.  Crystals of both materials are black and shiny, with \four\ forming as parallelogram platelets (see Fig.~\ref{fig:NPD}d inset) and \five\ (Fig.~\ref{fig:NPD}c) growing as hexagons.  


Growths of Mo-containing materials proceeded by a similar approach, using higher Al:Mo ratios between 125:1 and 150:1, and Mo powder from Aladdin (99.9\%).  Alumina lids were used on all Al--Mo growths.  The Al--Mo phase diagram is significantly more complicated than that of Al--W, with a cascade of noncentrosymmetric phases around the composition range between Al$_4$Mo and Al$_5$Mo\cite{Schuster1991,Saunders1997,Eumann2006,Peng2016,Cupid2010,Du2009}.  Most of the recent phase diagrams indicate an Al$_4$Mo phase that melts peritectically around 1150\,$^\circ$C and decomposes around 950\,$^\circ$C, Al$_{17}$Mo$_4$ which melts peritectically around 1000\,$^\circ$C, \ttMo\ with a peritectic decomposition at 950\,$^\circ$C but possibly only stable above 830\,$^\circ$C, then around 850\,$^\circ$C Al$_5$Mo begins to form.  Al$_5$Mo exhibits polymorphism, with the lowest-temperature structure being centrosymmetric $R\overline{3}c$\cite{Schuster1991,Eumann2006}.

A high-temperature Al--Mo growth, intended to produce Al$_{17}$Mo$_4$, was cooled from 995 to 945\,$^\circ$C over the course of several days, before free cooling to room temperature.  A growth intended to produce \ttMo\ was cooled from 935 to 855\,$^\circ$C over several days then cooled freely to room temperature.  An additional growth aiming for a high-temperature polymorph of Al$_5$Mo was initially heated to 900\,$^\circ$C to ensure full melting, then cooled in one hour to 730\,$^\circ$C, from whence it was cooled to 720\,$^\circ$C over the course of several days, then cooled at 200\,$^\circ$C/h to room temperature.  None of the Al--Mo growths produced the desired phase.

The Al--Mo crystals have a silvery metallic lustre.  The high-temperature Al--Mo crystals formed as large, thin hexagonal platelets (Fig.~\ref{fig:XRD}d), while those at slightly lower temperatures exist as much thicker hexagons (Fig.~\ref{fig:XRD}e).  The low-temperature growth produced flower-like clusters of tiny (100-300\,$\mu$m) hexagonal platelets which proved too small for resistivity measurements(Fig.~\ref{fig:XRD}c).

Resistivity was measured in zero field using a Quantum Design Physical Property Measurement System (PPMS) as a cryostat, but using an external lock-in detector.  A separate current source was used and an additional resistance of several kilohms was added in series with the current leads, for measurement stability.  Data were collected on cooling, then again on warming from base temperature to $\sim$20-50\,K to address an issue with rapid cooling rates in this temperature range.  For extracting a low-temperature power law, an offset was subtracted then a log-log plot was used to identify an appropriate upper temperature limit, and finally a least-squares fit to the function $\rho(T) = \rho_0 + AT^\alpha$ was performed to the original, unsubtracted data below that temperature.  Field-cooled and zero-field-cooled magnetization data were measured on all samples between 1.8 and 15\,K and in-plane fields in a PPMS using the vibrating sample magnetometry (VSM) option. Mosaics of crystals were mounted to a quartz bar with GE Varnish.  



Low-temperature specific heat was measured on mosaics of crystals between 0.1 and 4\,K using a PPMS with the dilution refrigerator option (tungsten materials) or the $^3$He refrigerator option (molybdenum materials).  Crystals were mounted using Apiezon N grease, with the short axis aligned along the magnetic field direction.  This corresponds to the monoclinic axis of \four, the hexagonal axis of \five, the [111] axis of the low-temperature Al--Mo material, and the $a$ axis of the higher-temperature Al--Mo crystals.  To enable more direct comparisons among the materials, the specific heat is calculated per mole of the transition metal atom.

Powder neutron diffraction data were collected at room temperature on the ECHIDNA diffractometer at the OPAL research reactor at ANSTO, Australia, from 4 to 164\,$^\circ$ in steps of 0.05\,$^\circ$, with a neutron wavelength of 1.6215\,\AA.  Single-crystalline samples of \four\ and \five\ were ground to powder, then spun to reduce the effect of preferred orientations.  Diffraction data were Rietveld-refined in {\scshape fullprof} by the least-squares method\cite{FullProf}.  Initial x-ray diffraction (XRD) was performed on single crystals and powders ground from crystals in a Bruker D8 Discover diffractometer, in powder geometry.  Single-crystal XRD was performed on a 0.21$\times$0.15$\times$0.08\,mm$^3$ crystal from the highest-temperature Al--Mo growth using a Bruker D8 Venture diffractometer with an APEX-II CCD area detector and a molybdenum K$\alpha$ source.  The structure was solved and refined using the {\scshape shelx} suite of software\cite{Sheldrick2015}.  Stoichiometries were verified by electron-probe microanalysis (EPMA), using a Shimadzu EPMA-1720, with a beam current of 10\,nA accelerated at 15\,kV.  Standard samples were the pure elements, using the Al K$\alpha$, W M$\alpha$, and Mo L$\alpha$ lines analyzed using a RAP (rubidium acid phthalate) crystal for Al, and a PET (pentaerythritol) crystal for the transition metals.

\section{Crystal Structures}

The original crystal structure refinements of \four\cite{Bland1958} and \five\cite{Adam1955} were performed in the 1950s using laboratory x-ray diffraction, and to our knowledge these have not been revisited.  Other materials crystallizing in the \four\ structure are now known to be prone to Al deficiency, and the atomic positions of \five\ have not been refined.  To address this, we begin by presenting crystal structure refinements of the tungsten materials based on neutron powder diffraction.

\begin{table*}[htbp]    
\caption{\label{tab:Al4W_NPD}Refinement of neutron powder diffraction data on  
  Al$_4$W in space group $Cm$ (No.\ 8) at 300\,K, with $a=5.25968(14)$\,\AA,
  $b=17.77333(45)$\,\AA, $c=5.22865(15)$\,\AA, $\beta=100.1088(10)^\circ$, and 
  $Z=6$: $R=5.56\%$, $R_f=4.20\%$, and $\chi^2=1.40$.  The refined composition is
  \three.}
\begin{tabular}{cclllll}\hline
Site & Mult.& $x/a$ & $y/b$ & $z/c$ & $U_{iso}$ (\AA$^2$) & Occ.\\ \hline \hline
W1 & $2a$ & 0 & 0 & 0 & 0.00197(47) & 1 \\
W2 & $4b$ & 0.3367(13) & 0.13745(21) & 0.3409(15) & 0.00197(47) & 1 \\
Al1 & $2a$ & 0.1437(28) & 0 & 0.5149(23) & 0.00497(62) & 1 \\
Al2 & $2a$ & 0.5037(25) & 0 & 0.1604(21) & 0.00497(62) & 1 \\
Al3 & $4b$ & 0.6593(19) & 0.07622(34) & 0.6903(18) & 0.00497(62) & 0.944(19) \\
Al4 & $4b$ & 0.8150(22) & 0.11802(37) & 0.2261(18) & 0.00497(62) & 0.952(25) \\
Al5 & $4b$ & 0.1862(25) & 0.12372(37) & 0.8127(20) & 0.00497(62) & 0.959(29) \\
Al6 & $4b$ & 0.6876(21) & 0.23514(36) & 0.5879(18) & 0.00497(62) & 0.904(22) \\
Al7 & $4b$ & 0.0241(20) & 0.25054(35) & 0.0761(22) & 0.00497(62) & 0.954(26) \\ \hline
\end{tabular}
\end{table*}

The crystal structure of \five, shown in Fig.~\ref{fig:NPD}b, is built up from Al$_{12}$ dodecahedral cages which closely resemble cubes with truncated corners, although the seemingly-flat square sides are not quite flat.  These dodecahedra are edge-sharing within the $ab$-plane and corner-sharing along the $c$-axis, and each cage contains one W atom.  

\four, shown in Fig.~\ref{fig:NPD}(a), has a more complex structure composed of layers of decahedral (Al$_{10}$) cages with hendecahedral (Al$_{11}$) cages above and below along the $b$-axis.  The W1 tungsten site is inside the decahedral cage, while W2 occupies the hendecahedra.  Each decahedron together with the hendecahedra above and below it form a face-sharing trimer, where the shared faces are quadrilaterals.  Connections between these trimers are edge-sharing along $b$ and corner- and edge-sharing within the $ac$-plane.  



The results of a structure refinement of \four\ are shown in Tab.\ \ref{tab:Al4W_NPD} and Fig.~\ref{fig:NPD}(d).  As expected based on other transition metal aluminides in the \four\ structure, an aluminum deficiency was observed.  The refined composition is \three, but this compound most likely inhabits a stability region with a temperature-dependent width.  EPMA offered further support for an aluminum deficiency, with an aluminum content of 3.53(36).  Several sites refined to full occupancy -- these occupancies were not refined further.  

\begin{table*}[htbp]
\caption{\label{tab:Al5W_NPD}Refinement of neutron powder diffraction
  data on Al$_5$W in space group $P6_3$ (No.\ 173) at 300\,K, with
  $a=4.98601(14)$\,\AA, $c=8.84923(20)$\,\AA, and $Z=2$: $R=4.92\%$,
  $R_f=6.35\%$, and $\chi^2=2.18$.}
\begin{tabular}{cclllll}\hline
Site & Mult.& $x/a$ & $y/b$ & $z/c$ & $U_{iso}$ (\AA$^2$) & Occ.\\ \hline \hline
W1 & $2b$ & 0.33333 & 0.66667 & 0.50000 & 0.00029(57) & 1 \\
Al1 & $2a$ & 0 & 0 & 0 & 0.00308(34) & 1 \\
Al2 & $2b$ & 0.33333 & 0.66667 & 0 & 0.00308(34) & 0.928(14) \\
Al3 & $6c$ & 0.3381(34) & 0.3394(31) & 0.24240(52) & 0.00308(34) & 1 \\ \hline
\end{tabular}
\end{table*}

Table \ref{tab:Al5W_NPD} and Fig.~\ref{fig:NPD}(e) report a refinement of \five.  Given the Al deficiency commonly reported in the \four\ structure, we refined the Al site occupancies for \five\ as well.  Only the Al2 site deviated from full occupancy.  The refined composition is Al$_{4.928(14)}$W.  EPMA returned aluminum contents of 4.68(16) and 4.95(15) in two batches of crystals, also suggestive of a nonstoichiometry.  An asymmetric peak shape in neutron diffraction in addition to the instrumental asymmetry, particularly at high angles, suggests a stability region with temperature-dependent width --- this would lead to a slight drift in lattice parameter over the course of the crystal growth.


\begin{figure}[htb]
  \includegraphics[width=\columnwidth]{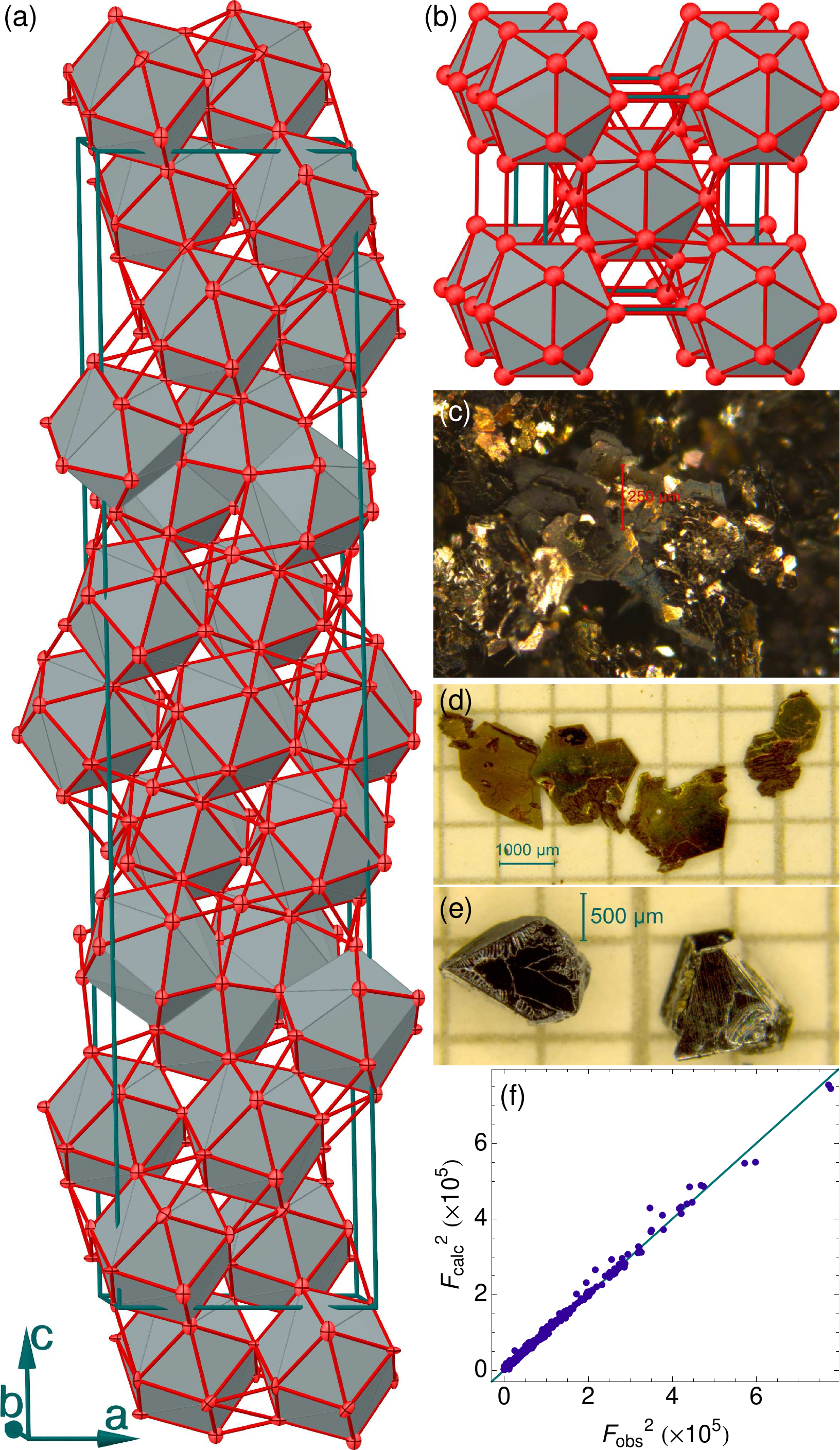}
  \caption{\label{fig:XRD}Structure of the molybdenum aluminides.  (a) Crystal structure of \fnMo, with the atoms shown as ellipsoids based on the refinement.  (b) Reported structure of \twelve\cite{Walford1964}.  (c) Crystals of \twelve, in the form of clusters of hexagonal platelets.  Crystals of \fnMo\ grown at (d) high and (e) lower temperatures.  (f) Demonstration of the quality of the crystal structure refinement of \fnMo.}
\end{figure}

Thorough investigations of the Al--Mo phase diagram have determined the crystal structures of all reported phases\cite{Schuster1991,Grin1995}, although the atomic positions and occupancies have generally not been refined.  We thus also investigated the structures of our Al--Mo crystals.  Our highest-temperature growth was found to be \fnMo, a new crystal structure with stacking very similar to that of \ttMo\ but with slightly different stoichiometry and different lattice parameters.  This material forms in the noncentrosymmetric monoclinic space group $C2$ (No.\ 5), with $a$=9.172(2)\,\AA, $b$=4.9393(13)\,\AA, $c$=41.072(11)\,\AA, and $\beta$=91.807(4)$^\circ$, as shown in Fig.~\ref{fig:XRD}a.  The details of the crystal structure refinement are presented in Tab.~\ref{tab:XRD1}, the refined atomic positions in Tab.~\ref{tab:XRD2}, and the refined anisotropic displacement parameters in Tab.~\ref{tab:XRD3}. The crystals grown at a slightly lower temperature had a lattice parameter consistent with the \fnMo\ structure and inconsistent with other known structures at higher Al content, but had poorer rocking curves that prevented single crystal structure refinement.  EPMA returned compositions of Al$_{4.56(5)}$Mo and Al$_{4.45(10)}$Mo for the higher- and lower-temperature growths, respectively, strongly suggestive of \fnMo\ = Al$_{4.\overline{45}}$Mo.  The physical properties of crystals grown under both sets of conditions were nearly identical, and it would be impossible to form the less-Al-rich \ttMo\ and Al$_{17}$Mo$_{4}$ with \fnMo\ separating them from the composition of the melt, so we conclude that the lower-temperature crystals are also \fnMo.  \fiveMo\ was not found.  The lowest-temperature crystals also had poor rocking curves, but diffraction off their large flat faces was consistent with the [111] axis of \twelve\cite{Adam1954,Walford1964}, the structure of which is shown in Fig.~\ref{fig:XRD}b.  Their formation in clusters of tiny plates may be a consequence of rapid crystallization at the conclusion of the growth.

\begin{table}[htbp]
  \caption{\label{tab:XRD2}Refined atomic positions for \fnMo\ in $C2$ (No.\ 5) at 203(2)\,K, with $a$=9.172(2)\,\AA, $b$=4.9393(13)\,\AA, $c$=41.072(11)\,\AA, $\beta$=91.807(4)$^\circ$, and $Z$=2.}
  \begin{tabular}{lcr@{.}lr@{.}lr@{.}lr@{.}l}\\ \hline\hline
    Site & Mult.& \multicolumn{2}{c}{$x$} & \multicolumn{2}{c}{$y$} & \multicolumn{2}{c}{$z$} & \multicolumn{2}{c}{$U_{eq}$ (\AA$^2$)}\\ \hline
    Mo1 & $4c$ & 0&18484(9) & 0&1094(3) & 0&04949(2) & 0&0046(2)\\
    Mo2 & $2b$ & \multicolumn{2}{l}{0} & 0&8577(3) & 0&5 & 0&0046(3)\\
    Mo3 & $4c$ & 0&05505(9) & 0&6079(2) & 0&14833(2) & 0&0047(2)\\
    Mo4 & $4c$ & 0&84178(9) & 0&3572(3) & 0&40127(2) & 0&0046(2)\\
    Mo5 & $4c$ & 0&92609(9) & 0&0944(2) & 0&24596(2) & 0&0044(2)\\
    Mo6 & $4c$ & 0&68455(9) & 0&8440(2) & 0&30365(2) & 0&0046(2)\\
    Al1 & $4c$ & 0&7974(3) & 0&8650(10) & 0&36605(8) & 0&0063(6)\\
    Al2 & $4c$ & 0&0212(3) & 0&1174(9) & 0&18370(8) & 0&0071(6)\\
    Al3 & $4c$ & 0&9559(3) & 0&3597(10) & 0&46618(8) & 0&0066(6)\\
    Al4 & $4c$ & 0&1506(3) & 0&6129(10) & 0&08381(8) & 0&0072(6)\\
    Al5 & $4c$ & 0&2194(3) & 0&6114(10) & 0&01606(8) & 0&0077(6)\\
    Al6 & $4c$ & 0&8845(3) & 0&8575(10) & 0&43407(8) & 0&0072(6)\\
    Al7 & $4c$ & 0&0890(3) & 0&1092(10) & 0&11602(8) & 0&0075(6)\\
    Al8 & $4c$ & 0&9552(3) & 0&6074(9) & 0&21442(8) & 0&0068(6)\\
    Al9 & $4c$ & 0&9251(3) & 0&5665(9) & 0&27994(8) & 0&0070(7)\\
    Al10 & $4c$ & 0&7230(3) & 0&3559(10) & 0&33540(8) & 0&0075(6)\\
    Al11 & $4c$ & 0&6734(3) & 0&3156(9) & 0&26981(8) & 0&0060(7)\\
    Al12 & $2a$ & \multicolumn{2}{l}{0} & 0&2644(10) & \multicolumn{2}{l}{0} & 0&0058(9)\\
    Al13 & $4c$ & 0&6688(3) & 0&5132(8) & 0&44993(8) & 0&0064(7)\\
    Al14 & $4c$ & 0&5036(3) & 0&0173(7) & 0&34898(8) & 0&0070(7)\\
    Al15 & $4c$ & 0&8677(3) & 0&7637(8) & 0&09985(8) & 0&0067(7)\\
    Al16 & $4c$ & 0&6911(3) & 0&7957(7) & 0&23776(8) & 0&0061(7)\\
    Al17 & $4c$ & 0&4517(3) & 0&5454(7) & 0&31210(8) & 0&0068(7)\\
    Al18 & $4c$ & 0&7320(3) & 0&2679(7) & 0&20070(9) & 0&0070(7)\\
    Al19 & $4c$ & 0&7267(3) & 0&0313(8) & 0&48446(8) & 0&0064(7)\\
    Al20 & $4c$ & 0&5661(3) & 0&5319(8) & 0&38431(8) & 0&0073(7)\\
    Al21 & $4c$ & 0&3244(3) & 0&7816(7) & 0&13493(8) & 0&0063(7)\\
    Al22 & $4c$ & 0&9544(3) & 0&7831(7) & 0&03455(8) & 0&0052(7)\\
    Al23 & $4c$ & 0&6146(3) & 0&0310(8) & 0&41496(8) & 0&0076(7)\\
    Al24 & $4c$ & 0&7844(3) & 0&7835(7) & 0&16551(8) & 0&0063(7)\\
    Al25 & $4c$ & 0&9149(3) & 0&2821(7) & 0&06548(8) & 0&0057(7)\\ \hline\hline
  \end{tabular}
\end{table}

The role of \fnMo\ in the Al--Mo phase diagram remains unclear and warrants further investigation.  The strong similarity in stacking between \ttMo\ and \fnMo\ suggests that these phases may be closely related.  To assist further research on the Al--W and Al--Mo systems, Crystallographic Information Files (CIF) describing the refinements are provided as Supplementary Materials (described in Appendix \ref{supp-CIF} and included in the ArXiv source).  

\section{Physical Properties}

Magnetization measurements (see Supplementary Materials, Appendix \ref{supp-M}) failed to identify any magnetic transition above 1.8\,K in any of the materials --- all were slightly diamagnetic, although a few had very weak low-temperature upturns suggestive of dilute magnetic impurities, most likely on the surface.

\begin{figure*}[htb]
  \includegraphics[width=\textwidth]{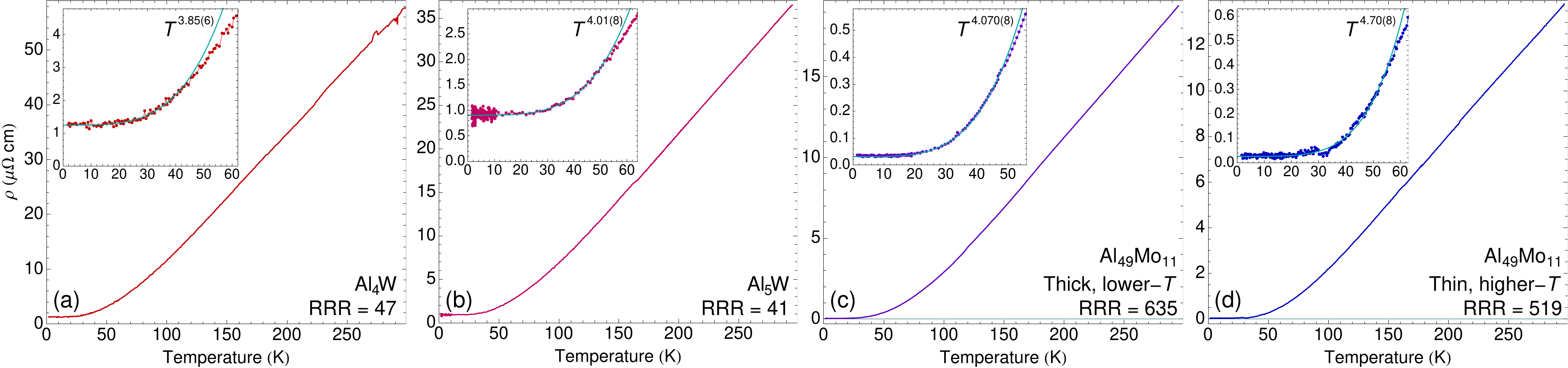}
  \caption{\label{fig:rho}Resistivity of the transition metal aluminides as labeled.  Their residual resistivity ratios (RRR) are listed, and insets demonstrate the respective low-temperature power laws.  \twelve\ was not measured due to the small size of the crystals.}
\end{figure*}

The resistivity $\rho$ of all materials measured indicates excellent metallic behavior with no sign of superconductivity nor any other phase transition above 1.5\,K.  The residual resistivity ratios [RRR $\equiv\rho(300\,\text{K})/\rho(0\,\text{K})$] are $\sim$45 for the W materials and $>$500 for \fnMo\ (see Fig.~\ref{fig:rho}), suggesting high sample quality.  While the tungsten materials' RRRs are lower than those found for the Mo compounds, these numbers are still large --- the non-stoichiometry evidently does not lead to a high residual resistivity.  The insets in Fig.~\ref{fig:rho} show power-law fits describing the low-temperature behavior.  Below 40--50\,K, the resistivity of all the materials is at least quartic in temperature, rather than the $T^2$ associated with Fermi liquids (for the full fit function and a comparison against a $T^2$ fit over the same temperature range, see the Supplementary Materials in Appendix \ref{supp-R}).  Such a $T^4$ power law has previously been observed in several elements, most notably in silver\cite{Barber1975,Barnard1982} but also in some Al alloys\cite{Powell1960}, and has been attributed to an interplay of electron-electron, electron-phonon, electron-impurity, and electron-dislocation scattering\cite{Bergmann1980,Wiser1982}.  Electron-spin-wave scattering, proposed to explain such a power law in rare earths and their compounds\cite{Mackintosh1963,Arajs1964,Ali1984}, is excluded in the nonmagnetic aluminides.  It is very uncommon for a $T^4$ power law to extend as high as 40--50\,K or over more than a decade in temperature as observed here.  

Recent theoretical work on noncentrosymmetric metals within Fermi liquid theory has indicated that the relaxation time due to electron-electron scattering $\tau_{ee}$, and therefore also the resistivity $\rho$, is essentially temperature-independent\cite{Mineev2018}. This applies for temperatures small compared to the spin-orbit band splitting, which is often hundreds of Kelvin.  This prediction would effectively eliminate the $T^2$ contribution, making electron-electron interactions just another component of the residual resistivity.  Since the predicted constant term depends crucially on details of the band structure at the Fermi level, it is unclear how {\slshape large} the constant contribution would be, and in particular whether it could be small enough in these materials to explain our results.  Lending support to this new interpretation, a low-temperature power law of $T^3$ has been reported in the noncentrosymmetric superconductor --- and excellent metal --- $\alpha$-BiPd\cite{Joshi2011b}, while other data on the same material\cite{Peets2016} can be fit to $T^{3.24(8)}$ below 25\,K. However, such increased power laws are certainly not universal --- the noncentrosymmetric superconductor Re$_6$Hf, whose breaking of time-reversal symmetry\cite{Singh2014} suggests strong spin-orbit effects, has a power law near $T^2$\cite{Singh2016}, while two noncentrosymmetric superconductors with non-Pauli-limited upper critical fields, (Nb,Ta)Rh$_2$B$_2$, have nearly-temperature-independent normal-state resistivity\cite{Carnicom2018}.  Further theoretical exploration in this area is clearly called for, particularly to clarify the effect of disorder, to identify any consequences that could be observed in other physical properties such as the specific heat, and to determine what would be expected in multi-band systems when not all Fermi surfaces have significant spin-orbit coupling.


\begin{figure}[htb]
  \includegraphics[width=0.7\columnwidth]{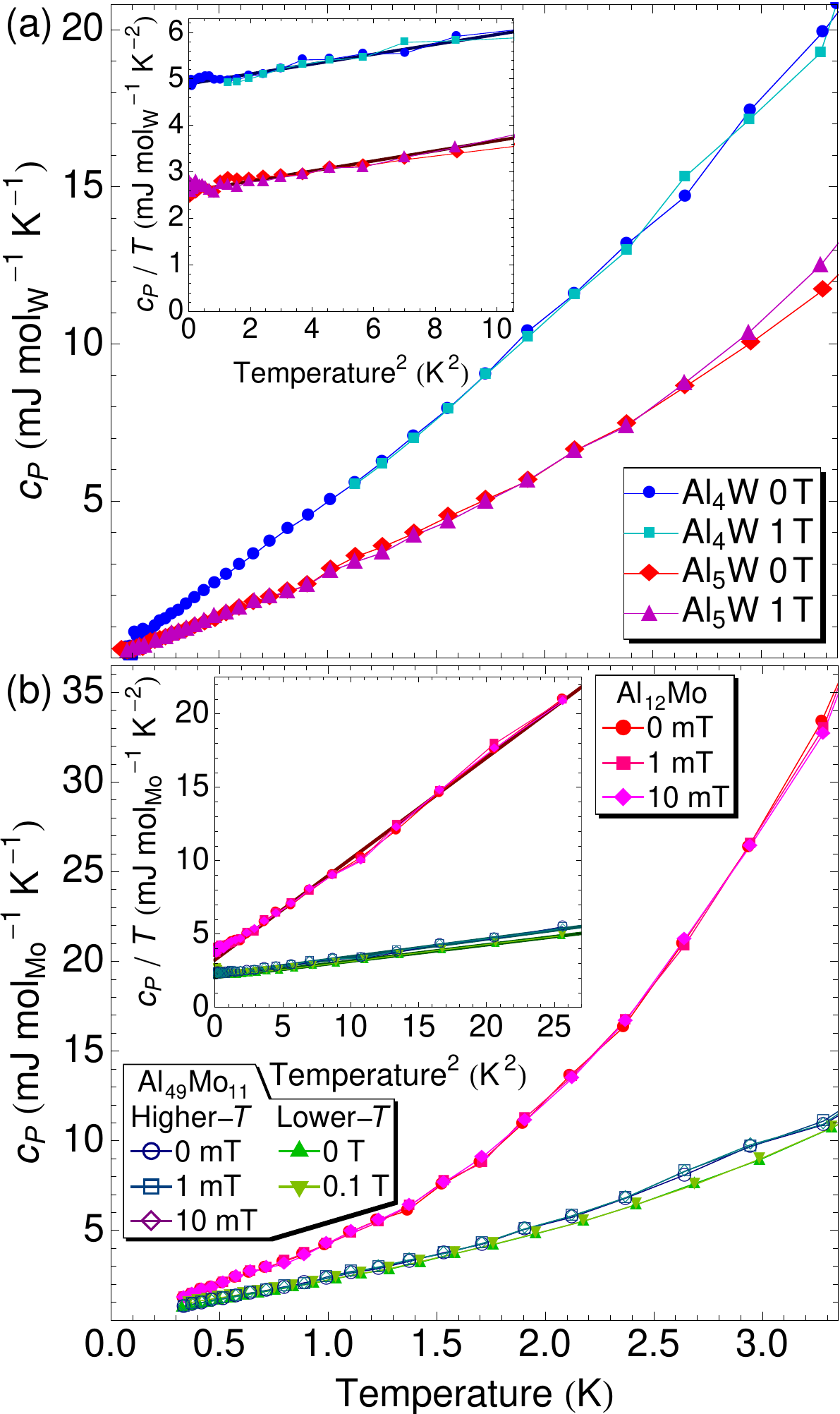}
  \caption{\label{fig:cP}Low-temperature specific heat of (a) \four\ and \five, in zero field and 1\,T, and of (b) \fnMo\ and \twelve.  The insets plot $c_P/T$ {\itshape vs}.\ $T^2$ and include best fit lines.  No phase transitions are observed.}
\end{figure}

Specific heat measurements on the tungsten aluminides [Fig.~\ref{fig:cP}(a)] found that neither \four\ nor \five\ has a thermodynamic phase transition above 100\,mK that would indicate superconductivity.  The Sommerfeld coefficients $\gamma$ representing the electronic specific heat contributions in \four\ and \five\ are 4.89 and 2.58\,mJ\,mol$^{-1}_\text{W}$\,K$^{-2}$, respectively, indicating that both compounds are conventional metals.  Their Debye temperatures $\Theta_D$ are 444 and 476\,K, respectively, in the low-temperature limit.  Results on the molybdenum aluminides [Fig.~\ref{fig:cP}(b)] are similar, placing a $\sim$300\,mK upper limit on phase transitions.  The Sommerfeld coefficients are 2.03 and 2.22\,mJ\,mol$^{-1}_\text{Mo}$\,K$^{-2}$, respectively, for mid- and high-temperature \fnMo, while their respective low-temperature Debye temperatures are 457 and 444\,K. \twelve\ has a Sommerfeld coefficient of 3.22\,mJ\,mol$^{-1}_\text{Mo}$\,K$^{-2}$ and a Debye temperature of 332\,K.  The minor differences in values between mid- and high-temperature \fnMo\ can likely be explained by contributions from a thin chloride film on the surface.  


Table~\ref{tab:all} summarizes the physical properties and crystal structure of the materials studied.

\begin{table}[htb]
  \caption{\label{tab:all}Summary of the physical properties and crystal lattice of the aluminides studied.  Molar quantities are calculated per mole of transition metal atom.  Density is calculated from diffraction results, at room temperature for the tungsten materials and at 203(3)\,K for \fnMo.  Structural properties of \twelve\ are taken from Ref.~\onlinecite{Walford1964}, and were determined at room temperature.}
  \begin{tabular}{lllll}\\ \hline\hline
    & \four & \five & \fnMo & \twelve \\ \hline
    RRR & 47 & 41 & $\sim$600 & ---\\
    $\rho(T)\sim T^n$, $n$= & 3.85(6) & 4.01(8) & $>4$ & ---\\
    $\gamma$ (mJ/mol$_M$\,K$^{2}$) & 4.89 & 2.58 & $\sim$2.1 & 3.22\\
    $\Theta_D$ (K) & 444 & 476 & $\sim$450 & 332\\
    Space group & $Cm$ & $P6_3$ & $C2$ & $Im$\=3\\
    $Z$ & 6 & 2 & 2 & 2\\
    $a$ (\AA) & 5.25968(14) & 4.98601(14) & 9.172(2) & 7.5815\\
    $b$ (\AA) & 17.77333(45) & 4.98601(14) & 4.9393(13) & 7.5815\\
    $c$ (\AA) & 5.22865(15) & 8.84923(20) & 41.072(11) & 7.5815\\
    $\beta$ ($^\circ$) & 100.1088(10) &  & 91.807(4) & \\
    Density (g/cm$^{3}$) & 5.9338(21) & 5.5224(66) & 4.2457(15) & 3.2095\\ \hline
  \end{tabular}
\end{table}

\section{Discussion and Conclusion}

Of the various novel ground states that can be observed in noncentrosymmetric materials, all forms of magnetic order can be immediately excluded on the basis of the magnetization results, which is unsurprising given the tendency of the elements involved to be nonmagnetic.  Topologically-protected states can also be excluded, at least near the Fermi level:  Such states exhibit spin-momentum locking, in which the spin orientation is uniquely defined for any given momentum.  As a consequence, scattering to a different momentum requires changing the spin orientation.  This excludes most scattering channels, leading to strongly enhanced conductivity --- topological states tend to dominate the transport at low temperature and low field.  The interplay of scattering mechanisms required to produce $T^4$ resistivity is incompatible with the presence of such states near the Fermi level --- in particular the highest-power-law contribution, electron-phonon scattering ($T^5$), is nonmagnetic and cannot produce a spin flip.  

The \Tc\ of Al is 1.2\,K\cite{Shoenberg1940,Cochran1958}, while Mo superconducts at 0.915\,K\cite{Geballe1962}, $\alpha$-W superconducts at 15\,mK\cite{Gibson1964} and metastable $\beta$-W superconducts at 1-4\,K\cite{Bond1965,Basavaiah1968}.  In a very simplistic view, it is unsurprising that the magnetization of compounds combining these elements shows no superconducting signal above 1.8\,K.  However, in the specific heat, the absence of superconductivity above 100-300\,mK in high-quality single crystals that are predominantly Al is more of a surprise.  Structurally, these materials are also evidently cage compounds, and rattling modes of the atom within similar cages are linked to enhanced superconductivity in skutterudites\cite{Pfleiderer2009}, $\beta$-pyrochlores\cite{Hattori2010}, clathrates\cite{Takabatake2014}, and possibly the exotic heavy-fermion superconductor UBe$_{13}$\cite{Flouquet2005,Pfleiderer2009} --- a low enough rattling frequency can hybridize with acoustic phonons, leading to strong anharmonicity and enhancing electron-phonon coupling, increasing the effective mass, and decreasing mobility\cite{Slack1995,Sales2003,Hattori2010}.  In the aluminides, however, it appears that the cage is too small to leave the W atom underconstrained:  The covalent radii of Al and W are 1.21 and 1.62\,\AA, respectively, which sum to 2.83\,\AA, while the Al--W bond lengths in \four\ and \five\ vary between 2.512(10) and 2.876(11)\,\AA.  The Mo--Al bonds in \twelve are 2.724\,\AA, while those in \fnMo\ vary between 2.625 and 2.958\,\AA. In both cases the shortest bonds are Mo--Al, rather than Al--Al. Rattling modes and the ensuing enhancement of superconductivity are thus not expected in these materials.



Matthias's results predict a minimum \Tc\ around 6 electrons per transition metal atom, with maxima for 5 and 7\cite{Matthias1955,Collver1973,Geballe2015}.  W and Mo, with 6, are both maximally suboptimal.  However, this ignores Al.  Neglecting the $\gtrsim$80\,\%\ of the atoms in an intermetallic metal which make up its bonding framework and instead expecting the properties to be determined solely by the relatively dilute transition metal would seem unlikely to adequately model the physical properties.

Another possible answer lies in the observation of a similar $T^4$ power law in the resistivity of Cu\cite{Powell1959} and Ag\cite{Barber1975,Barnard1982}, among other noble metals\cite{Matula1979}.  These metals have not been found to superconduct, most likely because their electrons lack a sufficiently strong interaction with the lattice.  The competition of interactions that leads to $T^4$ resistivity in those metals may imply insufficient electron-phonon coupling to form Cooper pairs.  If the $T^4$ power law in the aluminides arises from such a competition and implies similarly weak interactions, superconductivity may be suppressed to extremely low temperatures.

Finally, our non-observation of superconductivity may be a consequence of the difficulty of forming a stable pairing state in a spin-split band structure.  If the spin splitting is indeed found to be large in these materials, and if they can ultimately be shown to superconduct, the superconductivity will likely have strong singlet-triplet mixing.
If these materials do superconduct at lower temperatures, there is a large family of related aluminides of Cr, Mo, W, Mn, Re, and even Tc, plus substitution series among them --- the tuning of $d$-shell occupancy and spin-orbit coupling strength could be investigated in detail, albeit at very low temperatures.

In summary, we have investigated the low-temperature physical properties of several noncentrosymmetric tungsten and molydenum aluminides, finding the materials to be excellent metals but not identifying any transition to superconductivity down to 100-300\,mK.  A $T^4$ resistivity can be explained by a competition of scattering mechanisms, which may suggest weak electron-phonon coupling, and the aluminum cages are found to be too small to allow the rattling modes that are known to enhance superconductivity.  An alternative picture is that the lack of a $T^2$ term in the resistivity arises from strong spin-splitting at the Fermi level, which also makes it difficult to form Cooper pairs and reduces the energy saved through their formation.  These aluminides are relatively straightforward to grow, and are just a small part of a larger family of noncentrosymmetric cage aluminides.  In particular, the analogs hosting smaller $3d$ transition metals might be more likely to exhibit rattling modes, although spin-orbit-coupling-derived band splitting would be weaker.  If superconductivity can be found at lower temperatures in this large family, there is an excellent opportunity to tune the underlying interaction strengths and investigate the role of spin-orbit coupling in detail.  

\begin{acknowledgments}
This work was supported by the National Key R\&D Program of the MOST
of China (Grant No.~2016YFA0300200), the Science Challenge Program of
China, and the National Natural Science Foundation of China (Project
No.~11650110428).  The authors thank Yue-Jian Lin of the Shanghai Key 
Laboratory of Molecular Catalysis and Innovative Materials for 
assistance with single crystal x-ray diffraction, structure solution, 
and refinement.  The authors acknowledge the support of the Bragg 
Institute, Australian Nuclear Science and Technology Organisation, in 
providing the neutron research facilities used in this work.  Some of 
the groundwork for this study was supported by the Institute for 
Basic Science (IBS) in Korea (IBS-R009-G1).  
\end{acknowledgments}

\appendix
\section{Crystal Structure Refinement Details}

\begin{table}[htb]
  \caption{\label{tab:XRD1}Refinement details for a 210$\times$150$\times$80\,$\mu$m$^3$ crystal of \fnMo\ at 203(2)\,K.}
  \begin{tabular}{ll}\\ \hline\hline
    Formula & Al$_{49}$Re$_{11}$\\
    Space group & $C2$ (No.\ 5)\\
    $a$ & 9.172(2)\,\AA\\
    $b$ & 4.9393(13)\,\AA\\
    $c$ & 41.072(11)\,\AA\\
    $\beta$ & 91.807(4)\,$^\circ$\\
    $Z$ & 2\\
    $F(000)$ & 2198\\
    $\theta$ range & 2.98 to 27.53$^\circ$\\
    Index ranges & $-11\leq h\leq 11$,\\
     & $-6\leq k\leq 5$,\\
     & $-52\leq l\leq 49$\\
    Total reflections & 3342\\
    Reflections $I>2\sigma(I)$ & 3183\\
    Parameters & 273\\
    Restraints & 49\\
    Flack parameter\cite{Flack1983} & 0.00(4)\\
    Flack based on & 1001 pairs\cite{Flack2013}\\
    Goodness of fit & 1.122\\
    $R$ factors, all data & $R_1$=2.91\%, $wR_2$=8.76\%\\
    $R$ factors, $I>2\sigma(I)$ & $R_1$=2.67\%, $wR_2$=7.76\%\\
    Absorption coefficient $\mu$ & 4.755\,mm$^{-1}$\\
    Extinction coefficient & 0.00115(9)\\ \hline\hline
  \end{tabular}
\end{table}

Details of the single crystal x-ray diffraction structure refinement of \fnMo\ at 203(2)\,K are presented in Tab.~\ref{tab:XRD1}.  The $C2$ space group is enantiomorphic, so a Flack parameter was refined\cite{Flack1983,Flack2013}.  This parameter is near zero if the correct enantiomorph has been chosen or near unity if the refinement settled on the incorrect enantiomorph, while values near 0.5 indicate twinning or a racemic mixture.  The refined Flack parameter is zero.  All atomic positions were refined with anisotropic displacement parameters, which are presented in Tab.~\ref{tab:XRD3}.  Further information on the refinements is available in the CIF files, provided in the ArXiv source as Supplementary Materials as described in Appendix \ref{supp-CIF}.

\begin{table*}[htb]
  \caption{\label{tab:XRD3}Refined anisotropic displacement parameters in \AA$^2$ for \fnMo\ in $C2$ (No.~5) at 203(2)\,K.}
  \begin{tabular}{lr@{.}lr@{.}lr@{.}lr@{.}lr@{.}lr@{.}l}\hline\hline
    Site & \multicolumn{2}{c}{$U_{11}$} & \multicolumn{2}{c}{$U_{22}$} & \multicolumn{2}{c}{$U_{33}$} & \multicolumn{2}{c}{$U_{12}$} & \multicolumn{2}{c}{$U_{13}$} & \multicolumn{2}{c}{$U_{23}$} \\ \hline
    Mo1  & 0&0051(4) & 0&0040(4) & 0&0048(4) & 0&0001(4) & 0&0008(3) & 0&0002(4)\\
    Mo2  & 0&0046(5) & 0&0042(6) & 0&0052(6) & \multicolumn{1}{r}{0}& & 0&0012(4) & \multicolumn{1}{r}{0}&\\
    Mo3  & 0&0049(4) & 0&0043(4) & 0&0051(4) & 0&0001(4) & 0&0009(3) & 0&0004(4)\\
    Mo4  & 0&0044(4) & 0&0043(4) & 0&0050(4) & $-$0&0001(4) & 0&0009(3) & 0&0001(4)\\
    Mo5  & 0&0046(4) & 0&0045(4) & 0&0042(4) & $-$0&0002(4) & 0&0006(3) & $-$0&0002(4)\\
    Mo6  & 0&0044(4) & 0&0040(4) & 0&0054(4) & 0&0003(4) & 0&0011(3) & 0&0003(4)\\
    Al1  & 0&0065(13) & 0&0067(16) & 0&0058(15) & 0&0013(16) & 0&0002(11) & 0&0005(15)\\
    Al2  & 0&0092(13) & 0&0056(16) & 0&0065(14) & 0&0011(16) & 0&0015(11) & 0&0005(15)\\
    Al3  & 0&0037(13) & 0&0080(16) & 0&0080(15) & 0&0000(16) & $-$0&0002(11) & 0&0004(17)\\
    Al4  & 0&0106(14) & 0&0062(16) & 0&0048(14) & 0&0008(17) & 0&0008(11) & $-$0&0008(16)\\
    Al5  & 0&0106(14) & 0&0057(16) & 0&0069(15) & $-$0&0011(17) & 0&0025(11) & $-$0&0013(17)\\
    Al6  & 0&0062(13) & 0&0066(15) & 0&0090(15) & $-$0&0006(17) & 0&0003(11) & $-$0&0011(17)\\
    Al7  & 0&0093(13) & 0&0061(16) & 0&0072(15) & 0&0004(17) & 0&0015(11) & 0&0008(17)\\
    Al8  & 0&0076(13) & 0&0062(15) & 0&0068(14) & $-$0&0009(16) & 0&0002(11) & 0&0015(16)\\
    Al9  & 0&0072(13) & 0&0078(19) & 0&0060(15) & 0&0006(14) & 0&0004(11) & 0&0001(15)\\
    Al10 & 0&0070(13) & 0&0065(15) & 0&0089(15) & $-$0&0004(16) & 0&0002(11) & 0&0000(17)\\
    Al11 & 0&0034(13) & 0&0068(18) & 0&0078(15) & 0&0002(14) & 0&0001(11) & 0&0015(15)\\
    Al12 & 0&0022(18) & 0&009(2) & 0&006(2) & \multicolumn{1}{r}{0}& & $-$0&0015(16) & \multicolumn{1}{r}{0}&\\
    Al13 & 0&0046(14) & 0&0086(18) & 0&0060(16) & $-$0&0004(12) & 0&0004(12) & $-$0&0006(12)\\
    Al14 & 0&0074(14) & 0&0081(18) & 0&0056(16) & 0&0002(12) & 0&0018(12) & $-$0&0023(12)\\
    Al15 & 0&0037(14) & 0&0089(18) & 0&0075(16) & 0&0005(12) & $-$0&0013(12) & $-$0&0003(12)\\
    Al16 & 0&0042(13) & 0&0064(18) & 0&0076(15) & $-$0&0016(12) & $-$0&0009(11) & 0&0004(13)\\
    Al17 & 0&0062(13) & 0&0091(19) & 0&0050(15) & $-$0&0020(13) & 0&0006(11) & $-$0&0005(13)\\
    Al18 & 0&0068(14) & 0&0066(17) & 0&0076(16) & $-$0&0003(12) & $-$0&0004(12) & $-$0&0010(12)\\
    Al19 & 0&0079(14) & 0&0053(17) & 0&0061(16) & $-$0&0004(12) & 0&0022(12) & $-$0&0015(12)\\
    Al20 & 0&0063(14) & 0&0084(19) & 0&0072(16) & $-$0&0001(13) & 0&0008(12) & $-$0&0014(13)\\
    Al21 & 0&0038(13) & 0&0072(18) & 0&0080(16) & $-$0&0019(12) & $-$0&0010(11) & $-$0&0004(12)\\
    Al22 & 0&0024(12) & 0&0072(16) & 0&0059(14) & $-$0&0004(11) & $-$0&0018(10) & 0&0002(11)\\
    Al23 & 0&0081(14) & 0&0073(18) & 0&0076(16) & 0&0014(13) & 0&0015(12) & $-$0&0009(13)\\
    Al24 & 0&0039(13) & 0&0069(18) & 0&0081(16) & 0&0009(12) & 0&0011(11) & 0&0007(12)\\
    Al25 & 0&0019(13) & 0&0088(19) & 0&0063(16) & 0&0013(12) & $-$0&0012(11) & 0&0008(12)\\ \hline\hline
  \end{tabular}
\end{table*}

\section*{Supplementary Materials}
\renewcommand{\thefigure}{S\arabic{figure}}
\renewcommand{\thesection}{S\arabic{section}}
\setcounter{section}{0}

\section{Magnetization Data}\label{supp-M}

\begin{figure}[htb]
  \includegraphics[width=\columnwidth]{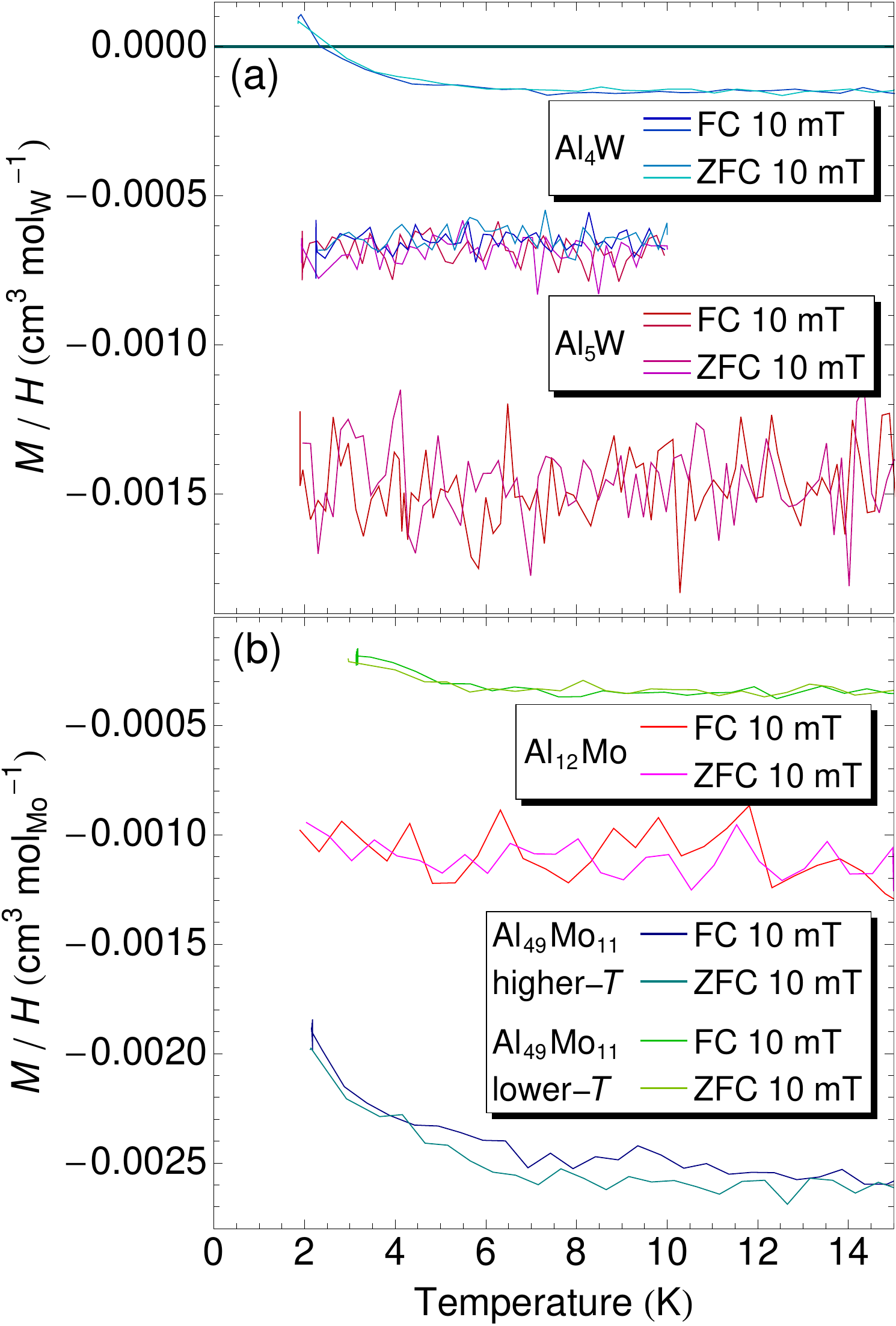}%
\caption{\label{fig:VSM}Magnetization of the (a) tungsten and (b) molybdenum aluinides in 10\,mT in-plane fields, under zero-field-cooled warming (ZFC) and field-cooled cooling (FC) conditions.  A paramagnetic contribution from the sample holder has been subtracted.  Variations between mosaics of the same material are most likely due to difficulties centering the sample due to its low signal, and contributions from chlorides on the surface.}
\end{figure}

Magnetization data on all materials are presented in Fig.~\ref{fig:VSM}.  Data were collected in a 100\,Oe in-plane field, under field-cooling and zero-field-cooling conditions.  All samples were diamagnetic before subtraction of a paramagnetic contribution from the sample holder, but the values are unlikely to be meaningful, primarily due to positioning issues --- the samples' signals were too weak to allow accurate centering.  Several of the samples exhibit a small extrinsic upturn at low temperature, which is most likely due to impurity phases on the surface.

\section{Resistivity Power Laws}\label{supp-R}

\begin{figure*}[htb]
  \includegraphics[width=\textwidth]{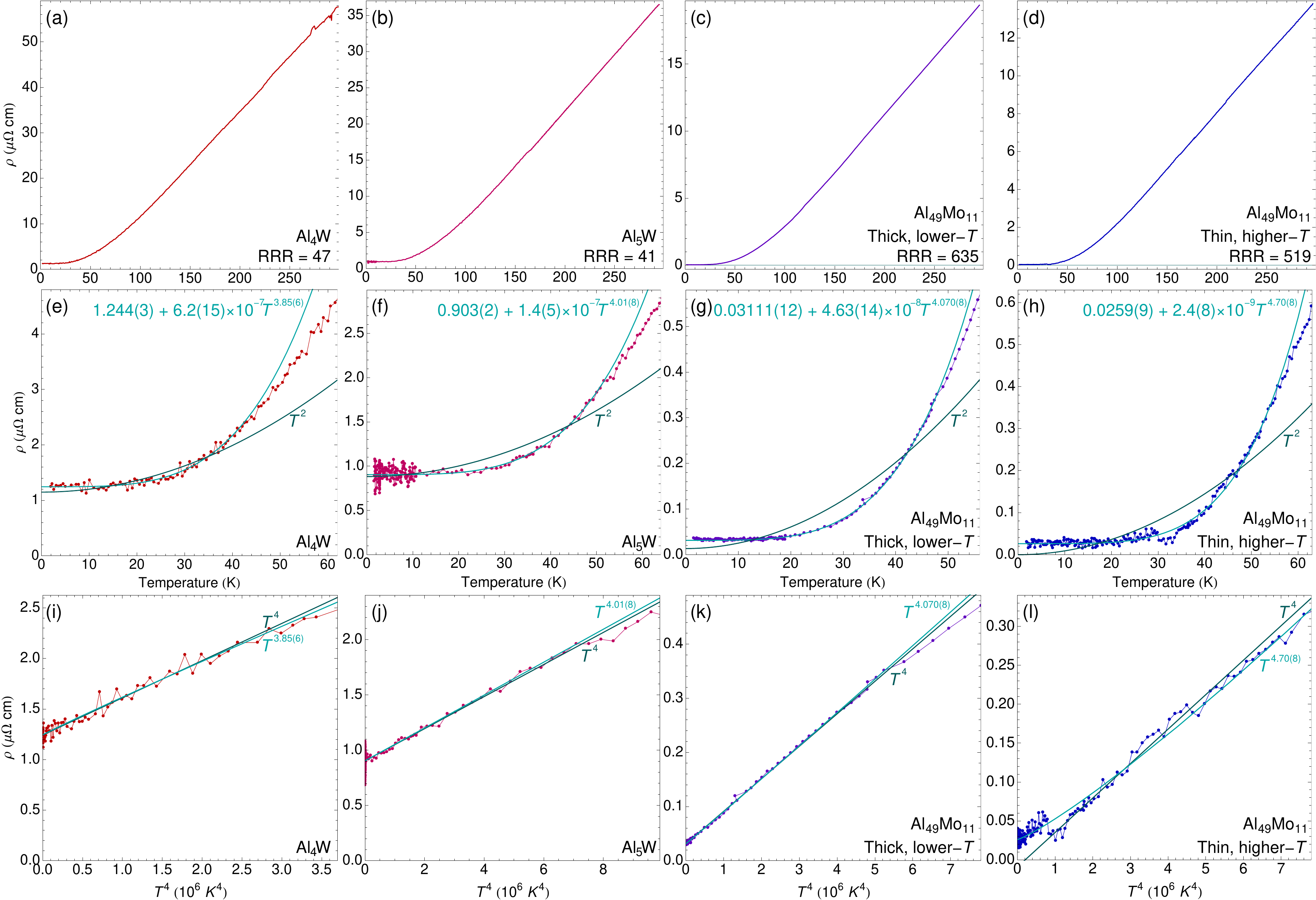}%
\caption{\label{fig:rhomore}Resistivity power laws.  Panels (a-d) are the resistivity as shown in the main text.  Panels (e-h) show the corresponding low-temperature power law fits, with the full fit functions included.  The constant intercept is in units of $\mu\Omega$\,cm and the power law's prefactor is in $\mu\Omega$\,cm\,K$^{-n}$, for the respective temperature power laws $T^n$.  Panels (i-l) compare the respective power law fits against a $T^4$ fit in linearized form.}
\end{figure*}

The resistivity data are replotted in Figs.~\ref{fig:rhomore}(a-d), with the full power law fit functions displayed.  A comparison against Fermi-liquid-like $T^2$ behaviour is provided in Figs.~\ref{fig:rhomore}(e-h) --- this form clearly does not adequately describe the data.  In Figs.~\ref{fig:rhomore}(i-l), the resistivity is plotted as a function of $T^4$, and a fit to $T^4$ is provided for comparison.  While the power law fits describe the data better, the deviations from $T^4$ are far less compelling than from $T^2$.  For each sample, all fits were performed over the same temperature range.  

\section{Diffraction Results}\label{supp-CIF}

Crystallographic Information Files (CIF) are provided for the three structure refinements, as separate files in the ArXiv source, to assist readers with structure visualization, modelling, and electronic structure calculations.

\bibliography{Al4W_arXiv}
\clearpage       
\end{document}